\documentclass[aps,prl,twocolumn,superscriptaddress,twoside, nofootinbib]{revtex4-1}
\usepackage{amsmath,amssymb,amsfonts}
\usepackage{graphicx}
\usepackage[english]{babel}
\usepackage{psfrag}
\usepackage{color}
\usepackage{tabularx}
\usepackage{times}
\usepackage{hyperref}
\usepackage{float}

\begin{document}

\title{Symmetry enhanced first-order phase transition in a two-dimensional quantum magnet}

\author{Bowen Zhao} 
\affiliation{Department of Physics, Boston University, 590 Commonwealth Avenue, Boston, Massachusetts 02215, USA}

\author{Phillip Weinberg} 
\affiliation{Department of Physics, Boston University, 590 Commonwealth Avenue, Boston, Massachusetts 02215, USA}

\author{Anders W. Sandvik} 
\email{sandvik@bu.edu} 
\affiliation{Department of Physics, Boston University, 590 Commonwealth Avenue, Boston, Massachusetts 02215, USA}
\affiliation{Beijing National Laboratory for Condensed Matter Physics and Institute of Physics, Chinese Academy of Sciences, Beijing 100190, China}

\date{\today}

\begin{abstract}
  Theoretical studies of quantum phase transitions have suggested critical points with higher symmetries than those of the underlying Hamiltonian. 
  Here we demonstrate a surprising emergent symmetry of the coexistence state at a strongly discontinuous phase transition between two ordered ground states. 
  We present a quantum Monte Carlo study of a two-dimensional $S=1/2$ quantum magnet hosting the antiferromagnetic (AFM) and plaquette-singlet 
  solid (PSS) states recently detected in {S\lowercase{r}C\lowercase{u}$_2$(BO$_3$)$_2$}. We observe that the O(3) symmetric AFM order and the Z$_2$
  symmetric PSS order form an O(4) vector at the transition. The control parameter $g$ (a coupling ratio) rotates the vector between the AFM and PSS 
  sectors and there are no energy barriers between the two at the transition point $g_c$. This phenomenon may be observable in 
  {S\lowercase{r}C\lowercase{u}$_2$(BO$_3$)$_2$}.
\end{abstract}

\maketitle

{\it Introduction.---}Theoretical studies of exotic  quantum states of matter and the phase transitions between them can provide new perspectives on many-body
physics and stimulate experimental investigations. A prominent example is the quantum phase transition between antiferromagnetic (AFM) and spontaneously dimerized
valence-bond solid (VBS) ground states in two-dimensional (2D) spin $S=1/2$ magnets \cite{Sachdev_NPH_2008,Kaul_ARC_2012}. Here the theory of deconfined quantum critical
points (DQCPs) suggests that the Landau-Ginzburg-Wilson (LGW) paradigm for phase transitions is inapplicable, as a consequence of quasi-particle fractionalization
\cite{Senthil_Sci_2004, Senthil_PRB_2004}. Over the past decade, likely DQCPs have been identified in lattice models, using ``designer Hamiltonians'' constructed
for their amenability to large-scale quantum Monte Carlo (QMC) simulations of the AFM--VBS transition
\cite{Sandvik_PRL_2007,Melko_PRL_2008,Jiang_JST_2008,Kuklov_PRL_2008,Lou_PRB_2009,Sandvik_PRL_2010,Chen_PRL_2013,Harada_PRB_2013,Pujari_PRL_2013,Block_PRL_2013,Nahum_PRX_2015,Shao_Sci_2016}.
Recently, a potential experimental realization of this type of DQCP was reported in the quasi-2D Shastry-Sutherland (SS) compound
{S\lowercase{r}C\lowercase{u}$_2$(BO$_3$)$_2$} under pressure \cite{Zayed_NP_2017}. Though the SS model \cite{Shastry_Phy_1981} is difficult to study numerically, due to
its geometrical frustration (which causes sign problems in QMC simulations), a specific type of VBS---a two-fold degenerate plaquette-singlet solid (PSS) located
between AFM and bond-singlet phases---was demonstrated convincingly by tensor-network calculations \cite{Corboz_PRB_2013}. Zayed et al.~\cite{Zayed_NP_2017} showed
that a PSS also exists in {S\lowercase{r}C\lowercase{u}$_2$(BO$_3$)$_2$} and suggested that the AFM--PSS transition  may be a DQCP. The phase transition was not
studied in the experiment, however, and it is not immediately clear if the two-fold degenerate PSS can support spinon deconfinement in the same way as a four-fold
degenerate VBS. QMC studies of rectangular lattices with two-fold degenerate VBS states point to a first-order transition \cite{Block_PRL_2013}, as was also found
in the SS model \cite{Corboz_PRB_2013}.

\begin{figure}[t]
\includegraphics[width=80mm,clip]{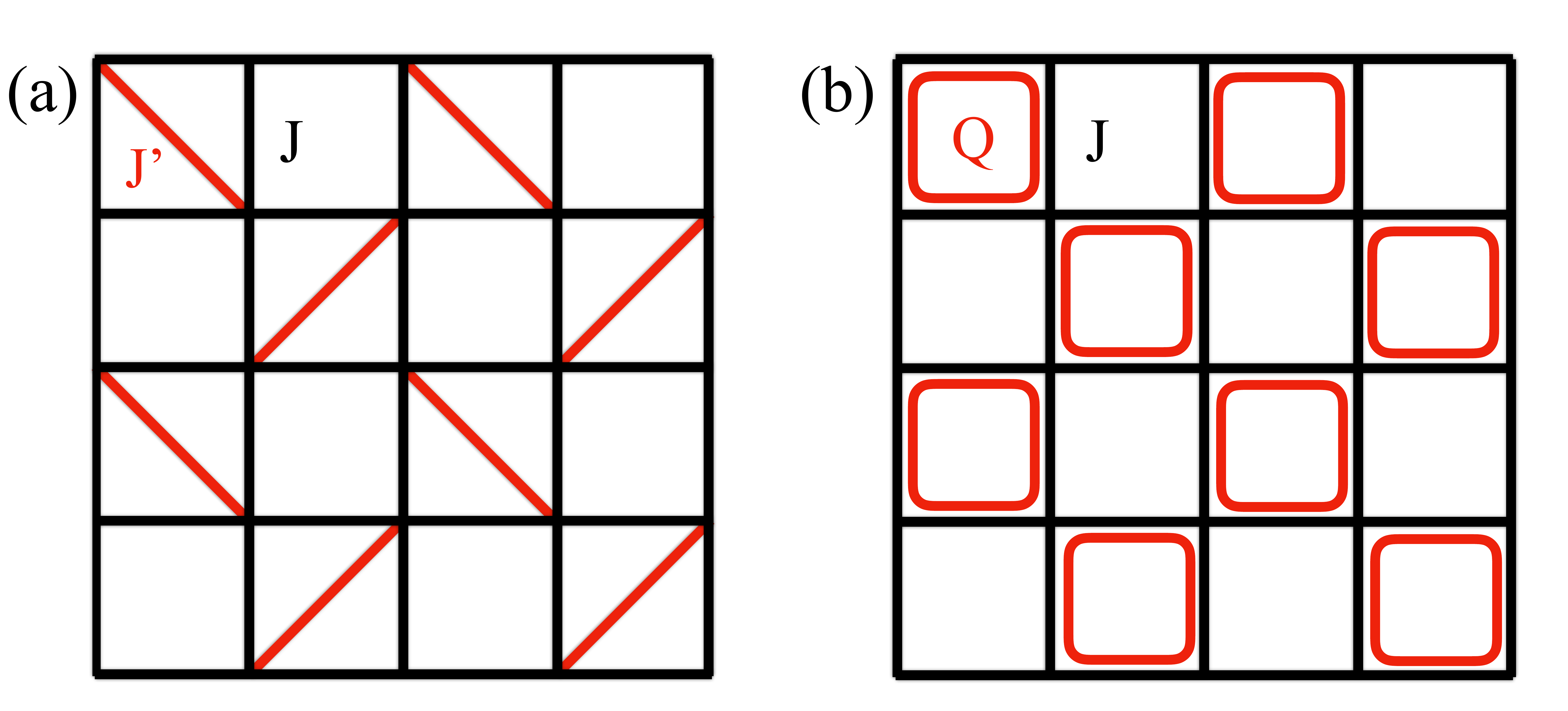}
\vskip-3mm
\caption{In the SS model (a), Heisenberg exchange $J$ between nearest neighbor $S=1/2$ spins
compete with next-nearest neighbor couplings indicated by diagonal lines. In the CBJQ 
model (b) the $J'$ interactions are replaced by the $Q$ terms in Eq.~(\ref{jqham}).}
\label{figmodels}
\end{figure}

Here we study a sign-free model that mimics the SS compound, in the sense that it shares the same kinds of AFM and PSS ground states.
The Hamiltonian, illustrated in Fig.~\ref{figmodels} along with the SS model, is a new member in the ``$J$-$Q$'' family
\cite{Sandvik_PRL_2007}, with Heisenberg exchange $J$ supplemented by four-spin interactions $Q$ that weaken and eventually destroy
the AFM order. Our QMC simulations demonstrate a first-order AFM--PSS transition with emergent O(4) symmetry.

Non-LGW critical points with emergent symmetries have been extensively investigated recently
\cite{Nahum_PRL_2015,Karch_PRX_2016,Metlitski_PRB_2016,Mross_PRL_2016,Kachru_PRL_2017,Wang_PRX_2017,Qin_PRX_2017,Sato_PRL_2017,Metlitski_2017,Gazit_2018,Stre_2018}.
In the case discussed here, the order parameters exhibit clear discontinuities but conventional phase coexistence is not observed. Using  order-parameter
distributions, we show that the AFM order is rotated by the control parameter into PSS order. Phase coexistence at the transition takes the form
of an O(4) symmetric vector arising out of the O(3) AFM and $Z_2$ PSS order parameters, with no energy barrier separating the two phases. In further
support of this scenario, we demonstrate a characteristic logarithmic form of the PSS ordering temperature versus the tuning parameter, as expected
for a 2D O($N\ge 3$) quantum system deformed by a $Z_2$ interaction \cite{Irkhin_PRB_1998,Cuccoli_PRB_2003}.

{\it  Ground states.---}Our Hamiltonian can be defined using singlet projection operators 
$P_{ij}=(1/4-{\bf S}_i \cdot {\bf S}_j)$;
\begin{equation}
H = -J \sum_{\langle ij \rangle} P_{ij} - Q \hskip-2mm \sum_{ijkl \in \Box^\prime} \hskip-1mm (P_{ij} P_{kl} + P_{ik} P_{jl}),
\label{jqham}
\end{equation}
where all indicated site pairs are nearest neighbors on a periodic square lattice with $L^2$ sites and $\Box^\prime$ denotes the $2\times 2$ $Q$-plaquettes
in Fig.~\ref{figmodels}(b). We define $g=J/Q$. For $g \to \infty$, this checker-board $J$-$Q$ (CBJQ) model reduces to the usual AFM ordered (at temperature
$T=0$) Heisenberg model, and for $g \to 0$ we will demonstrate a two-fold degenerate PSS. The model does not have any phase corresponding to the $J'$-bond
singlet state of the SS model for large $J'/J$. However, for elucidating the nature of the AFM--PSS transition, we can invoke symmetries and universality to propose
that the two models, as well as {S\lowercase{r}C\lowercase{u}$_2$(BO$_3$)$_2$}, contain the same physics.

We use two different QMC methods to study the CBJQ model: ground-state projection in the basis of valence bonds \cite{Sandvik_PRB_2010} and the stochastic
series expansion (SSE) method \cite{Sandvik_AIP_2010}. Both techniques deliver exact results to within statistical errors. The projector method is very useful
for studying spin-rotationally averaged quantities, while the SSE method is more efficient for finite-size scaling when the finite-$L$ ground states do not
have to be fully reached but $T \to 0$ as $L \to \infty$.

\begin{figure}[t]
\includegraphics[width=75mm,clip]{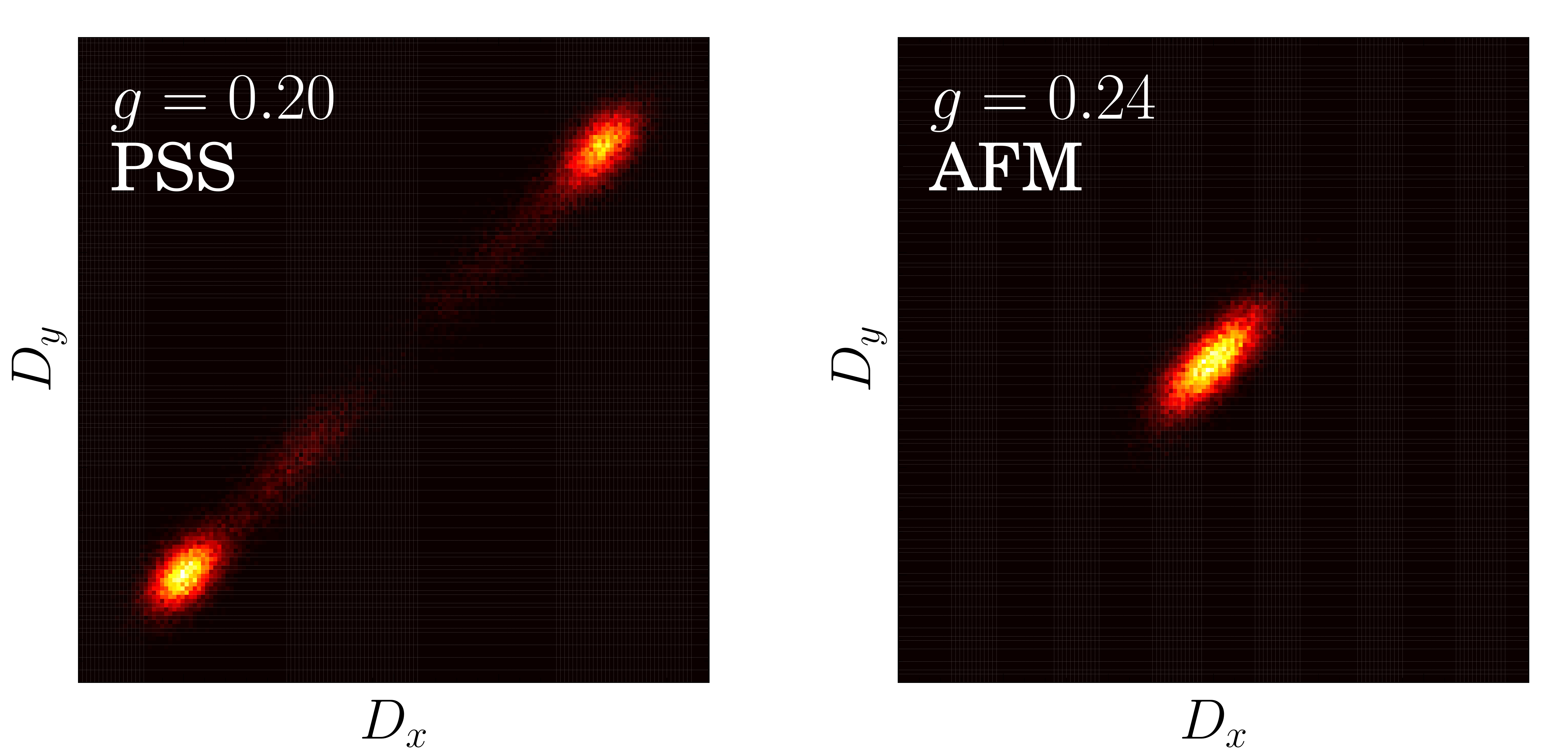}
\caption{Dimer order distribution $P(D_x,D_y)$ in the ground state of the CBJQ model at $g=0.20$ (in the PSS phase) 
and at $g=0.24$ (in the AFM phase), from valence-bond QMC on $L=96$ lattices.}\label{psdist}
\end{figure}

To demonstrate the PSS ground state for large $g$, we first study a conventional dimer order parameter
\begin{equation}
D_\mu = \frac{1}{L^2}\sum_{\mathbf{r}} (-1)^{r_\mu} {\bf S}(\mathbf{r})\cdot {\bf S}(\mathbf{r} + \hat{\mu}), \quad \mu = x, y,
\label{dxdydef}
\end{equation}
where the sum is over the lattice sites at $\mathbf{r}=(r_x,r_y)$. In a VBS, $\langle D_x\rangle \not =0,\langle D_y\rangle =0$ for $x$-oriented bond order
and the same with $x \leftrightarrow y$ for $y$-orientation. Since a singlet plaquette can be regarded as a resonance between horizontal and vertical bond
pairs, a two-fold degenerate PSS should have $|\langle D_x\rangle| = |\langle D_y\rangle| \not=0$ due to modulated singlet density on the plaquette rows
and columns in Fig.~\ref{figmodels}. On a finite lattice the symmetry is not broken, and the system fluctuates between the two states. We use the projector
method to generate the probability distribution $P(D_x,D_y)$. While strictly not a quantum mechanical observable, this distribution nevertheless properly
reflects the fluctuations and symmetries of the system. Results on either side of the AFM--PSS transition (the location of which will be determined below)
are shown in Fig.~\ref{psdist}. We see the two-fold symmetry of a PSS, instead of the four-fold symmetry of the columnar VBS
\cite{Lou_PRB_2009,Sandvik_PRB_2012}.

If the $Q$ terms are included on all plaquettes we arrive back to the original $J$-$Q$ model, whose AFM--VBS transition appears to be continuous
\cite{Shao_Sci_2016}. In accord with the DQCP theory, an emergent U(1) symmetry of its microscopically $Z_4$ invariant VBS order parameter has been
confirmed \cite{Sandvik_PRL_2007,Jiang_JST_2008,Sandvik_PRB_2012}. The proposed field theory description with spinons coupled to an U(1) gauge
field \cite{Senthil_Sci_2004, Senthil_PRB_2004} therefore seems viable. Unusual finite-size scaling behaviors not contained within the theory (but not
contradicted by it) have also been observed \cite{Sandvik_PRL_2010,Nahum_PRX_2015,Shao_Sci_2016} (and interpreted by some as a weak first-order
transition \cite{Kuklov_PRL_2008,Jiang_JST_2008,Chen_PRL_2013}). An interesting proposal is that the O(3) symmetry of the AFM and the emergent U(1)
symmetry of the VBS may combine into an SO(5) symmetry exactly at the critical point \cite{Senthil_PRB_2006,Nahum_PRL_2015}. In a spin-planar $J$-$Q$
model, it has instead been demonstrated that the U(1) AFM order parameter and the emergent U(1) VBS symmetry combine into a emergent O(4)
symmetry \cite{Qin_PRX_2017}. In yet another example, it was proposed that a system with O(3) AFM order and Z$_2$ Kekule VBS state exhibits a DQCP with
emergent SO(4) symmetry \cite{Sato_PRL_2017}. The O(3) and $Z_2$ symmetries apply also to the CBJQ model, and we therefore pay attention to a potential
O(4) or SO(4) symmetry \cite{so4note}.

{\it Finite-size scaling.---}To analyze the AFM--PSS transition, we perform SSE calculations at $T=2/L$. This way of taking the limit
$T \to 0, L \to \infty$ is appropriate for a quantum phase transition with dynamic exponent $z=1$, as well as a for
a first-order transition. We use order parameters defined solely with the $S^z$ spin components,
\begin{equation}
m_{z} = \frac{1}{L^{2}}\sum_{\mathbf{r}} \phi(\mathbf{r}) S^z(\mathbf{r}),~~
m_{p} = \frac{2}{L^{2}}\sum_{\mathbf{q}} \theta(\mathbf{q}) \Pi^z(\mathbf{q}),
\label{mssdef} 
\end{equation}
where the subscripts $z$ (spin component) and $p$ (plaquette) mark the AFM and PSS order parameters, respectively. In $m_z$, $\mathbf{r}$ runs over
all $L^2$ lattice sites and $\phi(\mathbf{r})=\pm 1$ is the staggered AFM sign. In $m_p$, we have defined an operator 
\begin{equation}
\Pi^z(\mathbf{q})=
S^z(\mathbf{q}) S^z(\mathbf{q} + \hat{x}) S^z(\mathbf{q} + \hat{y})S^z(\mathbf{q} + \hat{x} + \hat{y}),
\label{plaquettez}
\end{equation}
for detecting plaquette modulation, and the index $\mathbf{q}$ runs over the lower-left corners of the $Q$ plaquettes in Fig.~\ref{figmodels}. The
signs $\theta(\mathbf{q}) = \pm 1$ correspond to even or odd plaquette rows.

\begin{widetext}
  
\begin{figure}[t]
~~\includegraphics[width=178mm]{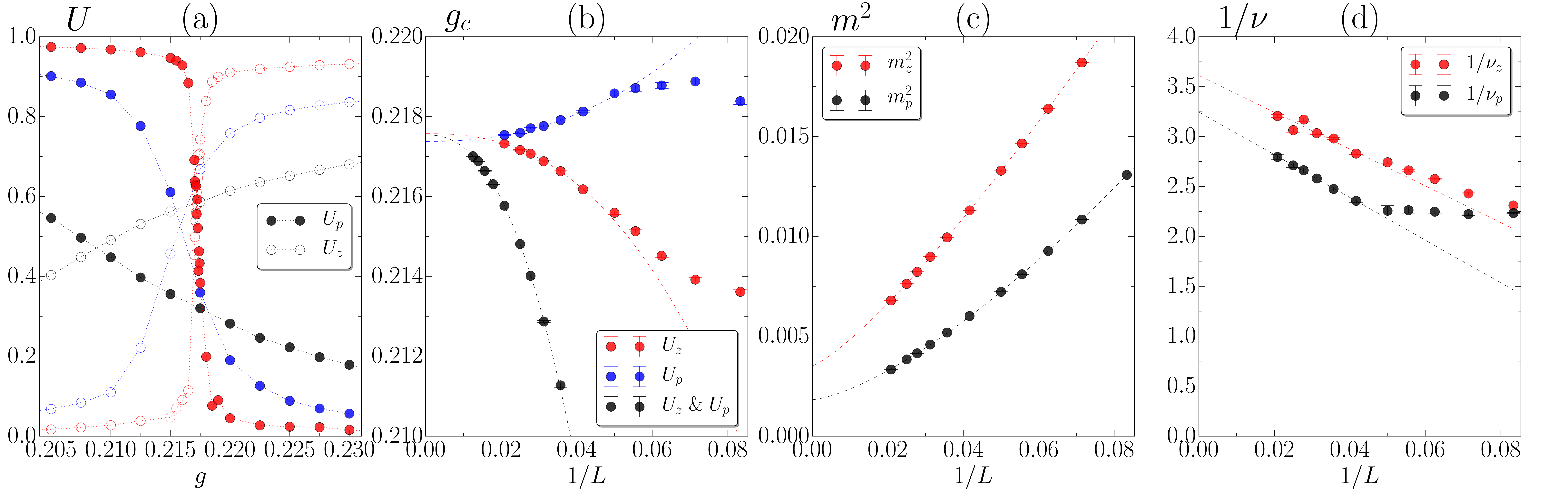}
\caption{Finite-size scaling of CBJQ results from SSE simulations at $T=2/L$. (a) Spin (open symbols) and plaquette (solid symbols) Binder cumulants versus
  $g$ for $L=24$ (black), $48$ (blue) and $96$ (red). Interpolations within these data sets (and results for other system sizes) underlie the analysis presented
  in the other panels. In (b) the crossing $g$-values of $U_z$ and $U_p$ are shown vs $1/L$ along with the $(L/2,L)$ same-quantity crossing points from $U_z$
  and $U_p$. The data extrapolate as $L\to \infty$ to $g_c=0.2175 \pm 0.0001$. The curves are fits including a single power-law correction $\propto L^{-\omega}$.
  In (c) the squared order parameters at the Binder $(L/2,L)$ cross points are graphed versus $1/L$ along with polynomial fits. The estimator of the
  correlation-length exponent, Eq.~(\ref{nudef}), is shown in (d) for both order parameters, along with line fits. Small system sizes were
  excluded from the fits until acceptable agreement with the functional forms were obtained.}
\label{jqresults}
\end{figure}

\end{widetext}

We will primarily analyze the Binder cumulants,
\begin{equation}
U_z = \frac{5}{2} \left(1 - \frac{\langle m_{z}^4\rangle}{3\langle m_{z}^2 \rangle^2}\right),~~~
U_p = \frac{3}{2} \left(1 - \frac{\langle m_{p}^4\rangle}{3\langle m_{p}^2 \rangle^2}\right),
\label{cumdef}
\end{equation}
shown in Fig.~\ref{jqresults}(a),
where the coefficients are chosen such that $U_z \to 1,U_p \to 0$ in the AFM phase while $U_z \to 0,U_p \to 1$ in
the PSS. If there is a single transition, we can use the crossing point $g=g^*(L)$ at which $U_z(g,L)=U_p(g,L)$ to define a finite-size critical point
$g^*(L)$. We also study the more commonly used crossing points of curves for two different system sizes, $bL$ and $L$ (where we use $b=1/2$), locating
the $g$ value where $U_z(g,bL)=U_z(g,L)$ or $U_p(g,bL)=U_p(g,L)$. The three definitions should flow to the same  $g_c$ when $L\to \infty$.

From the slopes of the cumulants we can extract the correlation-length exponents $\nu_z$ and $\nu_p$ \cite{Luck_PRB_1985,Shao_Sci_2016}: 
\begin{equation}
\frac{1}{\nu_{zp}} = \frac{1}{\ln(b)}\ln \left [\frac{dU_{zp}(g,L)/dg}{dU_{zp}(g,bL)/dg}\right]_{g=g_c(L)},
\label{nudef}
\end{equation}
where $g_c(L)$ is the relevant $(bL,L)$ cross point. The derivatives can be evaluated directly in the QMC simulations, and we interpolate to obtain the
cross points and slopes from data on a dense $g$-grid in the neighborhood of $g_c$.

The analysis is presented and explained in Fig.~\ref{jqresults}. We find
a single transition with $g_c=0.2175 \pm 0.0001$ based on all three cross point estimators in Fig.~\ref{jqresults}(b).  Most notably, in Fig.~\ref{jqresults}(c)
the order parameters at their respective Binder crossing points do not vanish as $L \to \infty$. This coexistence of AFM and PSS order is a decisive indicator
of a first-order transition. Another first-order indicator is seen in the exponents $1/\nu_z$ and $1/\nu_p$: At a classical first-order transition,
$1/\nu\to d$ in $d$ dimensions, and in 2+1 dimensions we might expect $1/\nu_{zp} \to 3$. The larger values seen in Fig.~\ref{jqresults}(d)
indicate a particular type of first-order transition, as explained below.

{\it Emergent O(4) symmetry.---}Due to energy barrier separating coexisting phases at a conventional first-order transition, the squared order
parameter follows a double-peaked distribution, which causes a divergent negative peak in the Binder cumulant
\cite{Vollmayr_ZPB_1993,Iino_ArX_2017}. Such peaks are present at the first-order transition in a $J$-$Q$ model with a staggered $Z_4$
VBS \cite{Sen_PRB_2010}, but are absent in Fig.~\ref{jqresults}(a). This lack of negative cumulant peaks leads us to consider alternative scenarios
for coexisting order parameters, without energy barriers. A well known case is a system with long-range order driven through a point at which the Hamiltonian
has a higher symmetry. As an example, we have studied an  XXZ-deformed 3D classical Heisenberg O(3) model in its ordered phase. As shown in detail in
Supplemental Material (SM) \cite{sm}, it behaves very similar to the CBJQ model if we make an analogy between the $xy$ magnetization and the AFM order
parameter on the one hand and the Ising magnetization and the PSS order parameter on the other hand.

The CBJQ model does not have any obvious enhanced symmetry, but the above results suggest that the O(3) AFM and the $Z_2$ PSS combine to form
an emergent O(4) symmetry at $g_c$ \cite{so4note}. In the transition region the system can then be
described by an effective deformed quantum O(4) model, where the control parameter $g = J/Q$ tunes the order parameter from the ordered $O(3)$ phase through the
O(4) point into the $Z_2$ phase.

As an explicit test of emergent O(4) symmetry, we use the valence-bond projector QMC method and now define the PSS order parameter with the rotationally
invariant operator 
\begin{eqnarray}
&&\Pi(\mathbf{q})={\bf S}(\mathbf{q})\cdot {\bf S}(\mathbf{q}+\hat x) + {\bf S}(\mathbf{q}+\hat y)\cdot {\bf S}(\mathbf{q}+\hat y + \hat x) \nonumber \\
&& ~~~+ {\bf S}(\mathbf{q})\cdot {\bf S}(\mathbf{q}+\hat y) + {\bf S}(\mathbf{q}+\hat x)\cdot {\bf S}(\mathbf{q}+ \hat x +\hat y ),
\label{pop}
\end{eqnarray}
in place of $\Pi^z(\mathbf{q})$ in Eq.~(\ref{mssdef}). For the AFM, we still use the $z$-component of the order parameter Eq.~(\ref{mssdef}).
In a state with both AFM and PSS order, the commutator $[m_z,m_p] \propto L^{-2}$, and we can safely use the c-numbers corresponding to $m_z$ and $m_p$
from a given transition graph \cite{Sandvik_PRB_2010} to accumulate the joint probability distribution $P(m_z,m_p)$. For the putative O(4) symmetry
to be manifest, we further normalize $m_z$ and $m_p$ by factors involving $\langle m_z^2\rangle$ and $\langle m_p^2\rangle$ \cite{sm}.

\begin{figure}
\includegraphics[width=84mm]{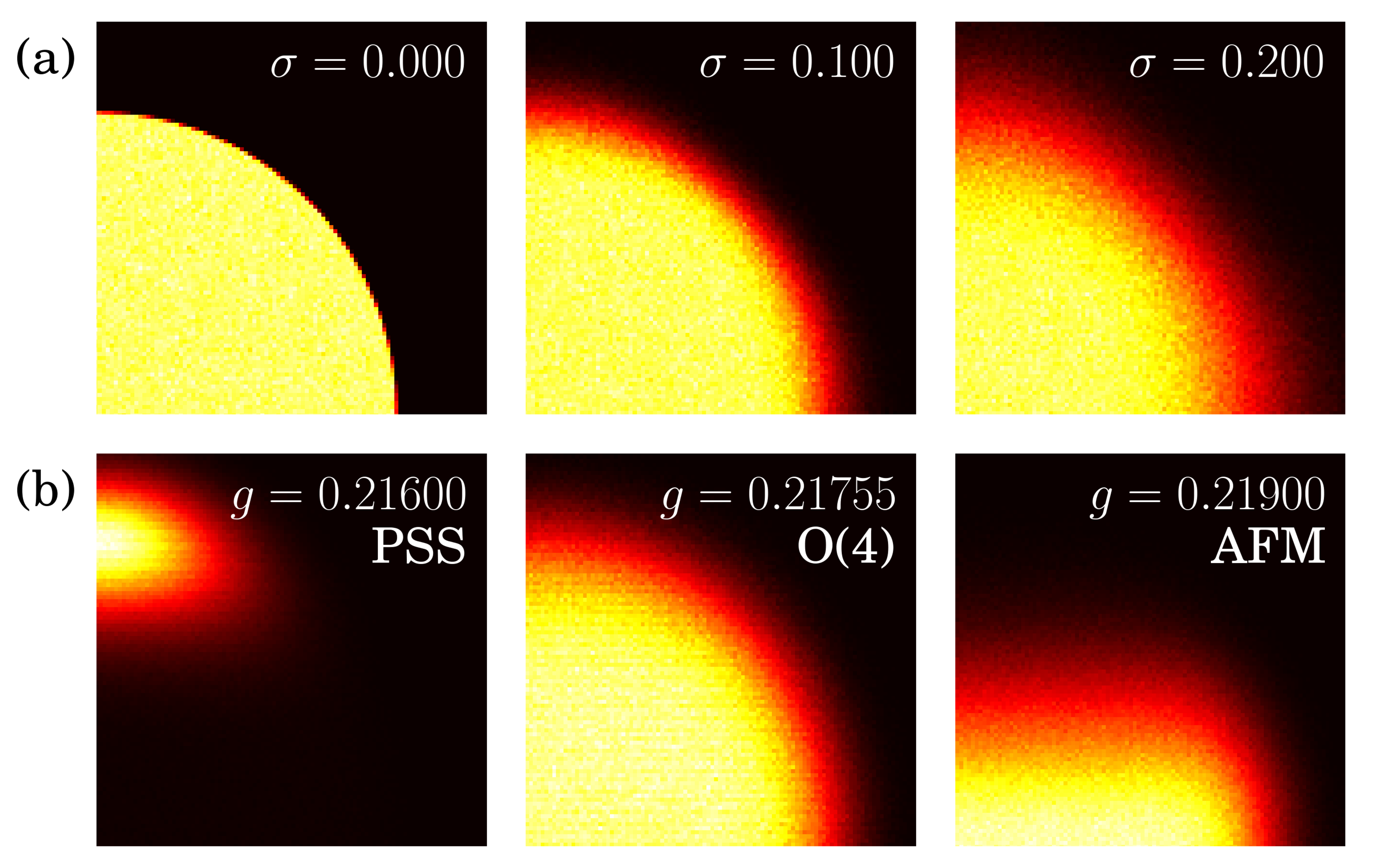}
\vskip-2mm
\caption{(a) One quadrant of the sampled \cite{Muller_ACM_1959} distribution of two components of an O(4) vector with Gaussian length fluctuations with
mean $R=1$ and standard deviation $\sigma$. (b) Projector QMC distribution $P(m_z,m_p)$ for the $L=96$ CBJQ model at three coupling ratios $g$. The $x$
axis represents the $z$ component of the AFM order parameter $m_z$, while the $y$-axis is the PSS order parameter $m_p$ \cite{sm}.}
\label{pmdist}
\end{figure}

For a point on an O(4) sphere of radius $R$, the projection onto two components results in a uniform distribution within a circle of radius $R$. However,
in a finite quantum system we also expect fluctuations of $R$ and therefore compare our CBJQ results with a distribution obtained from an O(4) sphere with
mean radius $R=1$ and standard deviation $\sigma$. Examples are shown Fig.~\ref{pmdist}. At the transition, the CBJQ distribution is rotation symmetric with
radial profile similar to O(4) sampling with $\sigma \approx 0.15$. Inside the phases the distributions are shifted as expected---deep in the PSS we should
eventually, for $L \to \infty$, obtain a point on the $y$-axis, and in the AFM state a line on the $x$-axis. Quantitative tests of the symmetry are 
presented in SM \cite{sm}. As expected for an emergent symmetry, we find clear O(4) violations for small system sizes ($L=8,16$), but no detectable
deviations at $g_c$ for the largest systems studied (up to $L=96$).

Having concluded that there is emergent O(4) symmetry, we can also understand why $1/\nu_{z,p} >3$ in Fig.~\ref{jqresults}(b): The dynamic exponent of
the Anderson-Goldstone rotor states associated with O($N\ge 3$) order is $z=2$, and therefore one may expect the exponents to eventually tend to $d+z=4$
when $L\to \infty$ at $T=0$. The deviations may be due to $T>0$ effects when $T$ is scaled as $L^{-1}$ (instead of $L^{-2}$). As we show in SM \cite{sm}, quantitative
measures of the emergent O(4) symmetry in our $T=0$ calculations exhibit $L^{-4}$ scaling of the size of the $g$-window in which the symmetry is emergent. 

Another interesting consequence of O(4) symmetry should be a specific logarithmic (log) form of the critical PSS temperature $T_c$ versus the distance
$\delta=g_c-g$ from the $T=0$ transition point, $T_{\rm c} \propto \log^{-1}(C / \delta)$, as in an O($N \ge 3$) model with an Ising deformation
\cite{Irkhin_PRB_1998,Cuccoli_PRB_2003}. This form is very different from that expected close to an Ising quantum-critical point, where
$T_{\rm c} \propto \delta^{\nu_{\text{3D}}}$, where $\nu_{\text{3D}}$ is the 3D Ising correlation-length exponent. Neither form should apply at a
conventional first-order transition extending from $(g_c,T=0)$ to some $T>0$. If the O(4) breaking perturbation is very weak, one should still
expect the log form to hold down to some low temperature.

\begin{figure}
\includegraphics[width=80mm]{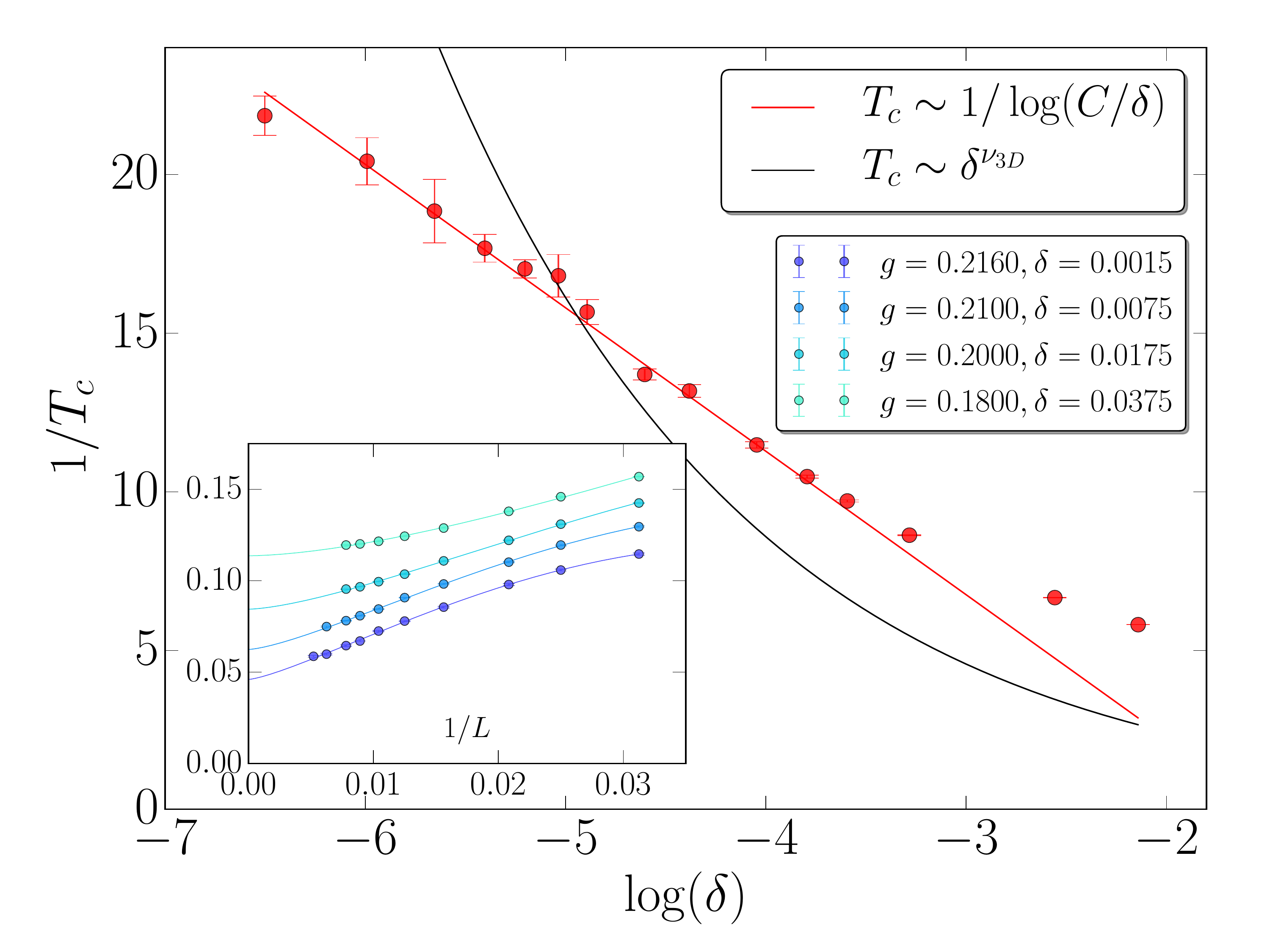}
\caption{Inverse PSS critical temperature versus the shifted coupling ratio $\delta=g_c-g$. The red line is a fit to the expected log form, and the black
curve is of the conventional Ising form as a contrast. The inset shows examples of the extrapolation of $T_c$ using the expected critical scaling form with a
subleading correction, $T_c = aL^{-b}(1+cL^{-d})$, with fitting parameters $a,b,c,d$ and $L$ up to $160$.}
\label{Fig:Tc}
\end{figure}

We have computed $T_c(g)$ for the PSS by the cumulant-crossing method using SSE data for $L \le 160$. We can reliably extrapolate $T_c$ to the thermodynamic
limit for $g \le 0.216$ ($\delta \agt 0.0015$), as shown in Fig.~\ref{Fig:Tc}. The behavior for $\delta \alt 0.02$ is very well described by the
log form, lending strong indirect support to the emergent O(4) symmetry through an important physical observable in the
thermodynamic limit. 

{\it Discussion.---}We cannot exclude that the O(4) symmetry is present only up to some length scale above the largest system, $L=96$, studied
here. Such symmetry violations at a long scale may be expected at certain weak first-order transitions, either when the system is close to a fine-tuned point
with order parameter of the higher symmetry (though no convincing emmergent symmetries were observed in connection with this scenario \cite{Kuklov04}), or
in proximity of a quantum-critical point at which the higher symmetry is emergent \cite{Nahum_PRL_2015,Wang_PRX_2017,Metlitski_2017}. In the latter case,
perturbations break the symmetry above some length scale $\xi'$ larger than the correlation length $\xi$ \cite{Wang_PRX_2017}.

In the CBJQ model studied here,
the observed discontinuities are rather strong; from Fig.~\ref{jqresults}(c), the magnitude of the O(4) vector in AFM units is
$m_s=\langle 4m_z^2\rangle^{1/2} \approx 0.12$, almost $25\%$ of the maximum  staggered magnetization $1/2$. The first-order nature of the transition
is apparent even on small lattices, e.g., as seen in the flow of $1/\nu_z$ toward an anomalously large value in Fig.~\ref{jqresults}(d). Thus, in the scenario
of Ref.~\cite{Wang_PRX_2017}, we should have $\xi \ll L \ll \xi' \sim \xi^{1+a}$, where the exponent $a$ must be rather large in order to give the clear
separation of length scales needed to account for the observed behavior. Such behavior has not been previously anticipated; rather, emergent symmetry on
large length scales has been cited as support for continuous non-LGW transitions \cite{Nahum_PRL_2015,Sato_PRL_2017}

In an alternative scenario of an asymptotic O(4) symmetry, the dominant symmetry-breaking field is tuned to zero at the AFM-PSS
transition and higher-order O(4) violating perturbations would vanish upon renormalization, perhaps by an extension of the DQCP
framework, or by some more general mechanism. While emergent O($N$) multicritical points arising from O($N-1$)
and $Z_2$ order parameters have been extensively discussed within the LGW framework \cite{Hebert_PRB_2001,Pelissetto_2007,Hasenbusch_2011,Eichorn_2013},
the influence of the higher symmetry on associated first-order lines have not been addressed until recently in the weakly first-order DQCP
context \cite{Wang_PRX_2017}. In order to exclude that the CBJQ model is accidentally fine-tuned to vanishing or extremely small perturbations
of the O(4) symmetry, we have also studied a model extended by additional interactions; see SM \cite{sm}.

We designed the CBJQ model with {S\lowercase{r}C\lowercase{u}$_2$(BO$_3$)$_2$} in mind. The expected Ising-type $T>0$ PSS
transition would be a good target for detecting the still incompletely characterized PSS phase and the putative O(4) symmetry. In 2D we have
demonstrated a characteristic log form of $T_c$ (Fig.~\ref{Fig:Tc}), and this form will hold down to some low temperature in the presence
of sufficiently weak inter-layer couplings; presumably the 3D $T=0$ transition is a conventional first-order one.

The O(4) AFM--PSS transition is reminicent of the SO(5) theory of high-$T_c$ superconductivity \cite{demler04}, where O(3) AFM and O(2) superconducting
order parameters form the higher symmetry; a scenario not confirmed, neither experimentally nor in models. After the completion of the
present work, an SO(5) analogue of the AFM--PSS was demonstrated in a spin-$1$ $J$-$Q$ model \cite{Kaul18}, and an O(4) transition very similar
to ours was discussed in the context of a classical 3D loop model \cite{Serna18}.

\begin{acknowledgments}
  {\it Acknowledgments.---}We would like to thank Fakher Assaad, Ribhu Kaul, Naoki Kawashima, Shiliang Li, Zi Yang Meng, Adam Nahum, Ying Ran, Subir Sachdev,
  Hui Shao, Liling Sun, and Zhi-Cheng Yang for stimulating discussions. This work was supported by the NSF under Grant No.~DMR-1710170 and by a Simons
  Investigator Award. The calculations were carried out on Boston University's Shared Computing Cluster.
\end{acknowledgments}

\setcounter{equation}{0}
\setcounter{figure}{0}
\renewcommand{\theequation}{S\arabic{equation}}
\renewcommand{\thefigure}{S\arabic{figure}}

\begin{widetext}

\begin{center}  

\null\newpage

\section{Supplemental Material}

{\bf \noindent Symmetry enhanced first-order phase transition in a two-dimensional quantum magnet}
\vskip2mm

{\noindent
Bowen Zhao, Phillip Weinberg, and Anders W. Sandvik}
\vskip2mm

\end{center}

We present additional results supporting the existence of an emergent O(4) symmetry in the CBJQ model at its AFM--PSS transition.
In Sec.~1, as a benchmark for finite-size scaling, we discusss results for the phase transition between an $xy$-ordered state and a $z$-ordered
state in a deformed classical O(3) model (the XXZ model) below its critical temperature. In Sec.~2 we carry out a quantitative analysis of the
order parameter histograms exemplified in Fig.~\ref{pmdist} of the main paper. In Sec.~3 we deform the CBJQ model by introducing alternating
plaquette terms of strength $Q_A$ and $Q_B$, such that the case $Q_A=Q,Q_B=0$ corresponds to the original CBJQ model discussed in the main text. We find
emergent O(4) symmetry also with $Q_B=Q_A/2$, showing that the O(4) symmetry of the original CBJQ model is not accidental due to some implicit
fine-tuning.
\vskip5mm

\end{widetext}

\subsection{1. Classical XXZ model}

In the 3D classical O($N$) model at $T<T_c$, a deformation of one of the $N$ interaction terms by a factor $\Delta$ leads to an ordered phase
breaking O($N-1$) symmetry for $\Delta<1$ and $Z_2$ (Ising) symmetry for $\Delta>1$. We have argued that the CBJQ model corresponds to this
situation with $N=4$, with the $O(3)$ and $Z_2$ phases corresponding to the AFM and PSS, respectively. However, in the CBJQ model the O(4)
symmetry is not explicitly present at the Hamiltonian level but is emergent on long length scales.

We will here study the deformed 3D classical O(3) model and demonstrate finite-size scaling behaviors analogous to those that we found for
the CBJQ model. We could also have studied an O(4) model, but the same physics is manifest already with $N=3$, which is the minimum
number of components for which one of the deformed phases breaks a continuous symmetry and the other one breaks the discrete $Z_2$ symmetry.
The Hamiltonian of the XXZ model is:
\begin{equation}
H = - \sum_{\langle i j \rangle}\sigma^x_i\sigma^x_j+\sigma^y_i\sigma^y_j+ \Delta\sigma^z_i\sigma^z_j,
\label{Eq:ClModel}
\end{equation}
where $\langle i j \rangle$ corresponds to nearest-neighbor interactions between the unit vectors $\mathbf{\sigma}_i$ on a simple cubic lattice. We
could also consider the 2D $S=1/2$ AFM Heisenberg model at $T=0$ with a similar deformation, which was done in Ref.~\onlinecite{Cuccoli_PRB_2003}
but with a different focus.

The 2D $S=1/2$ XXZ model with nearest-neighbor interactions on the square lattice has long-range order at $T=0$ for all values of
$\Delta$, and the order parameter symmetry changes with $\Delta$ in the same way as in the 3D classical model below $T_c$. When passing through
the special point $\Delta=1$, the elementary excitations change, as the Goldstone modes present in the O(2) phase and the O(3) point are gapped out
continuously for $\Delta > 1$. In this sense, we can consider the change in symmetry as a phase transition with both first-order and continuous
characteristics; a discontinuous flip of the direction of the order parameter but a continuously varying gap. Note that $T>0$ in the
quantum XXZ model corresponds to a finite size $L_3 \propto 1/T$ of one of the dimensions of the classical 3D model (here written as the third
dimension). Here we will only conside the 3D model with equal size $L$ in all dimensions and, thus, obtain results
corresponding to $T=0$ in the quantum case when $L\to \infty$.

We have carried out Monte Carlo simulations of the classical 3D XXZ model at two different temperatures in the ordered phase, $T^{-1}=0.7$ and $0.75$, the former
being close to $T^{-1}_c(\Delta_c=1)\approx 0.6930$. We use the efficient Wolff cluster agorithm \cite{wolff_alg} and analyze the $xy$ and $z$ magnetizations
individually. The Binder cumulants and slopes are defined in ways analogous to Eqs.~(\ref{cumdef}) and (\ref{nudef}).

As shown in Fig.~\ref{hbresults1}, behaviors very similar to those in the CBJQ model (Fig.~\ref{jqresults}) are observed at At $T = 0.7$ if we make an analogy
between the $xy$ magnetization and the AFM order parameter on the one hand and the Ising $z$ magnetization and the PSS order parameter on the other hand.
There are no negative cumulant peaks in Fig.~\ref{hbresults1}(a) in the neighborhood of the transition point $\Delta=1$. In Fig.~\ref{hbresults1}(b) we show
that the transition point is accurately reproduced with the same curve-crossing method as we used for the CBJQ model. Looking at the coexistence values of the
order parameters in Fig.~\ref{hbresults1}(c), which obey $\langle m^2_x\rangle + \langle m^2_y\rangle = 2\langle m^2_z\rangle$, the length $|m|$ of the vector ${\bf m}$
is about 25\% of the maximum value $1$. This is similar to the relative length of the O(4) vector extracted from Fig.~\ref{jqresults}(c) for the CBJQ model,
where we can define $|m|=\sqrt{4\langle m_z^2\rangle}$, with maximum value $|m|=1/2$. 

In Fig.~\ref{hbresults1}(d) we see that $1/\nu_{xy}$ extrapolates close
to the standard first-order value $3$, while $1/\nu_z$ is somewhat lower. One may question the standard first-order value in this case because of the lack
of free-energy barriers, but a simple mean-field argument gives that the relevant scaling variable is $(\Delta -1)L^3$ also for the long-range ordered
XXZ model. In Fig.~\ref{hbresults1}(d) we have only used a simple line fit to extrapolate both the exponents, and an asymptotically form with higher-order 
corrections and larger system sizes would likely explain why the values are less than $3$. Similar to the CBJQ model, Fig.~\ref{jqresults}(d),
there is a rather sharp cross-over 
from one slope at small system sizes to a higher slope for the larger sizes. The cross-over region should correspond to $L \approx \xi$, $\xi$ being the 
correlation length of the infinite system in the neighborhood of the transition. Because of the cross-over behavior, it is difficult to carry out reliable 
extrapolations with the rather small number of system sizes available.

In most respects, we see that the O(3) order--order transition looks in finite-size scaling like a first-order transition, with the glaring exception
of the lack of negative Binder peak. Indeed, with phase coexistence in the form of a higher symmetry, the arguments behind the negative peak
\cite{Vollmayr_ZPB_1993,Iino_ArX_2017} do not apply, since, thanks to the higher rotational symmetry, the two phases are not separated by a free-energy barrier
at the transition point $\Delta=1$. These results for the classical model provide support for a similar mechanism at play at the AFM--PSS transition of the
CBJQ model, even though no exact higher symmetry is present in its Hamiltonian.

In Fig.~\ref{hbresults2} we show results at a slightly lower temperature, $T^{-1}=0.75$. The results are qualitatively very similar to those at
$T^{-1}=0.7$, but the features are sharper due to the stronger order. The extrapolation of both exponents $1/\nu_{xy}$ and $1/\nu_z$ to $3$ is also
clearer in this case, with an overall weaker size dependence (though still the linear extrapolation somewhat underestimates the values of both the
exponents) and no clear cross-over. The lack of cross-over here suggests that the correlation length in the infinite system remains small at this
temperature.

\begin{widetext}
  
\begin{figure}[t]
\includegraphics[width=168mm]{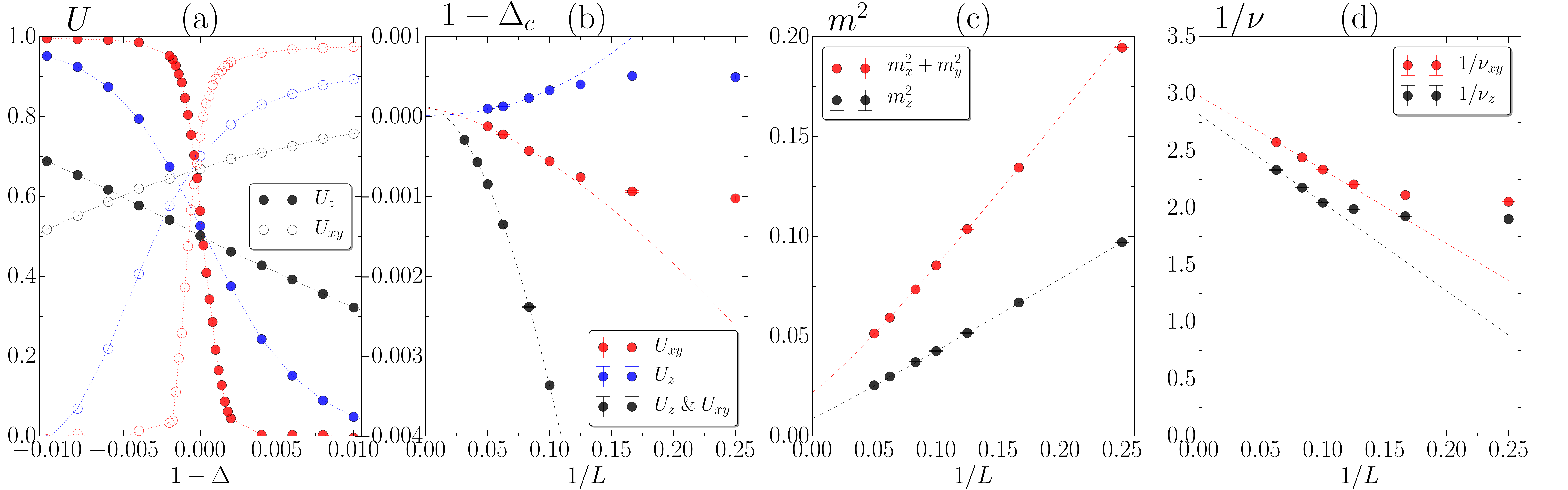}
\caption{Results for the classical 3D Heisenberg with anisotropy $\Delta$ at $T^{-1}=0.7$, graphed as in Fig.~\ref{jqresults} in the main paper;
the $xy$ and $z$ Binder cumulants vs $\Delta$ in (a), the crossing points extrapolated in $1/L$ to the phase transition point in (b), the coexistence values
of the squared order parameters versus $1/L$ in (c), and the correlation-length exponents extrated from the slopes of the cumulants in (d). The system sizes
in (a) are $L=8$ (black),$16$ (blue) and $32$ (red), with open and solid symbols used for $U_{xy}$ and $U_z$, respectively.}
\label{hbresults1}
\end{figure}

\begin{figure}[h]
\includegraphics[width=168mm]{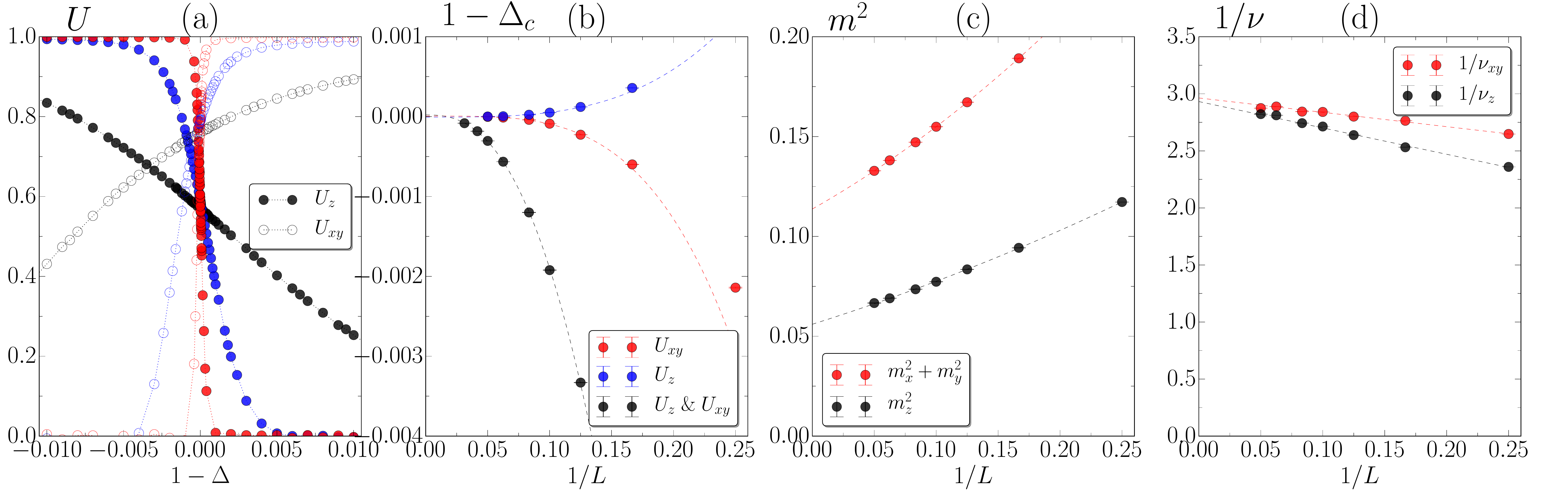}
\caption{The same quantities as in Fig.~\ref{hbresults1} for the 3D XXZ model at a lower temperature, $T^{-1}=0.75$.}
\label{hbresults2}
\end{figure}

\end{widetext}

\subsection{2. Quantitative check of emergent O(4) symmetry}
\label{o4analysis}

Here we discuss further details of our tests of emergent O(4) symmetry in the CBJQ model based on order-parameter distributions (histograms)
$P(m_z,m_p)$ such as those shown in Fig.~\ref{pmdist}. In addition, we also consider the distribution of $P(m_s,m_p)$, where $m_s$ is the magnitude
of the full $O(3)$ AFM order parameter,
\begin{equation}
m^2_s = m_x^2 + m_y^2 + m_z^2. 
\label{msdef}
\end{equation}
Like $m_p$ defined in Eqs.~(\ref{mssdef}) and (\ref{pop}), the c-number corresponding to $m_s^2$ is obtained in the valence-bond projector QMC method after each
Monte Carlo updating sweep directly from the transition graph as a single unique number (in contrast to just the component $m_z$, which is obtained by sampling
one of the many spin configurations that contribute to the transition graph) \cite{Sandvik_PRB_2010,Liang1988,Beach2006}.
Note that it is not possible to obtain independent
equal-time values for all three components of the AFM order parameter from the transition graphs or the associated $z$ basis spin configurations.

In the simulations, we generate and store a list of points $(m_z,m^2_s,m_p)_i$, $i=1,\ldots,M$. In order to obtain smooth probability distributions
and small error bars on the associated integrated quantities that we use to test for the emergent symmetry, we need a very large number of points ($M$ of
the order of millions). To capture the point of maximal symmetry and study the scaling properties away from this point, we also need many values of $g$. We
have carried out the valence-bond QMC simulations with a very long projection time $\tau$ in the imaginary-time evolution operator ${\rm e}^{-\tau H}$ and
find that $\tau=8L$ is sufficient for $T \to 0$ convergence of the distributions for the system sizes consiedered here; up to $L=96$.
\vskip2mm

{\it Symmetry tests with two components.}---The definitions of the two order parameters by Eqs.~(\ref{mssdef}) and (\ref{pop}) are not unique. Therefore,
even if there is an emergent symmetry between the order parameters, $m_z$ and $m_p$ are not directly comparable as to their overall magnitudes. To investigate
a possible emergent O(2) symmetry of the distribution $P(m_z,m_p)$, as a proxy for the full O(4) symmetry of all four components, we need to remove the ambiguity
by properly normalizing the sampled numbers. To this end, post-simulation, we compute the corresponding variances $\langle m_z^2\rangle$ and $\langle m_p^2\rangle$.
We can then define the radius $R$ of the distribution as
\begin{equation}
\langle R^2\rangle = \langle m_z^2\rangle + a^2\langle m_p^2\rangle,~~~~R \equiv \langle R^2\rangle^{1/2},
\end{equation}  
while also requiring that
\begin{equation}
\langle m_z^2\rangle = a^2\langle m_p^2\rangle.
\label{mcondition}
\end{equation}  
Thus, the parameter $a$ that puts the two sampled order parameters on an equal scale is defined by
\begin{equation}
a^2 = \frac{\langle m_z^2\rangle}{\langle m_p^2\rangle}.
\label{ascale}
\end{equation}  
We can now define normalized point pairs as
\begin{equation}
(\tilde m_z,\tilde m_p ) = R^{-1} (m_z,am_p),
\label{mmscaled}
\end{equation}  
and test for emergent O(2) symmetry in the distribution $P(\tilde m_z,\tilde m_p )$ at the AFM--PSS transition.

To quantify the degree of O(2) symmetry of a distribution $P(\tilde m_z,\tilde m_p )$ we use the integrals
\begin{eqnarray}
I_q &=& \int \text{d}\tilde{m}_z \text{d}\tilde{m}_p P(\tilde{m}_z, \tilde{m}_p) \cos \bigl (q \phi(\tilde{m}_z, \tilde{m}_p ) \bigr )\nonumber \\
& = &  \frac{1}{M}  \sum_{i=1}^{M}\cos(q \phi(|\tilde{m}_z|, |\tilde{m}_p|)_i),
\label{iqint}
\end{eqnarray}
where on the second line $i$ is the index corresponding to the $M$ QMC sampled points $({m}_z,{m}_p)_i$, from which angles
$\phi(|\tilde{m}_z|, |\tilde{m}_p|)_i \in [0,\pi/2]$ are extracted (with the absolute values taken to transform to the positive
quadrant). We will here consider the integrals with $q = 2, 4, 6, 8$, all of which should vanish if the distribution is O(2) symmetric.
For larger $q$, the results become increasingly noisy, but since there is no reason to expect distributions with $I_2=I_4=I_6=I_8=0$ and
$I_{q>8} \not=0$, what we do here is sufficient for demonstrating O(4) symmetry.

There is a remaining ambiguity in the normalization, as to the point at which the scale factor $a$ should be evaluated. In Fig.~\ref{pmdist}(b) of
the main paper, $a$ was evaluated at $g=0.21755$ (the data in the middle panel) and used at the other $g$ values as well. If the distribution is O(2)
symmetric at $g_c$, as it appears to be, we argue that the best way to proceed is to fix $a$ at this point, instead of using a $g$-dependent value $a^2(g)$
computed from a distribution that is not O(2) symmetric when $g \not= g_c$. This choise is motivated by the fact that the O(4) symmetry should only
apply to a single point, the transition point, and there is no reason why the normalization condition Eq.~(\ref{mcondition}) should be applied elsewhere.
If the varying $a^2(g)$ is used for all values of $g$, then the distribution away from $g_c$ will artificially be drawn out in one directon, thus making the
signals (the values of $I_q$) less sensitive to the control parameter. The fact that the mean vector projection in the different directions should not be
the same when the vector flips from the AFM sector to the PSS direction would then be missed. Note that fixing $a^2(L)$ at $g_c$ also in no
way can artificially introduce a false symmetry in the quantities we study.

\begin{figure}[t]
\includegraphics[width=84mm,clip]{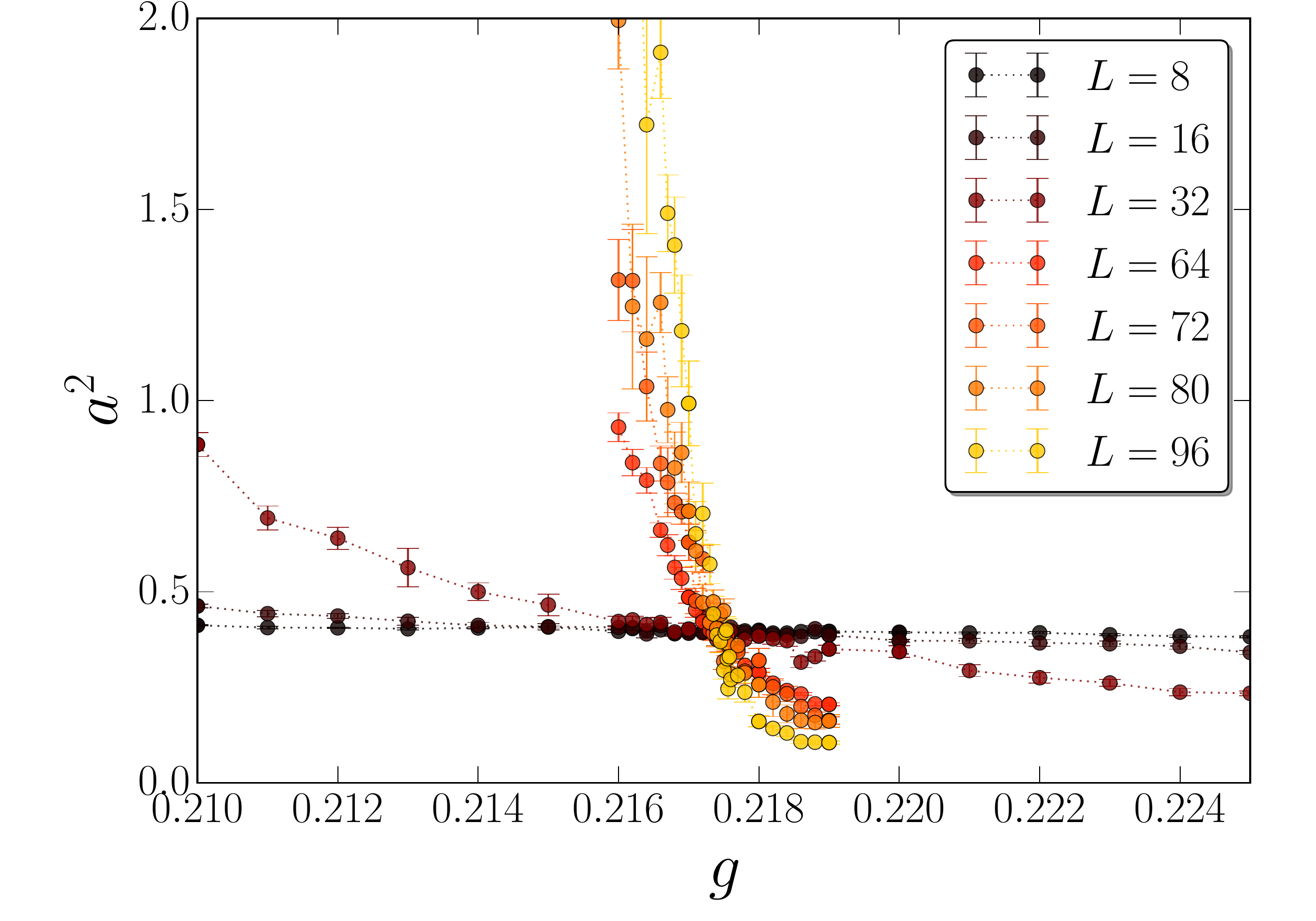}
\vskip-1mm
\caption{The dependence on the coupling $g$ and system size $L$ of the order-parameter normalization ratio $a^2$ defined in Eq.~(\ref{ascale}).}
\label{supfig1}
\end{figure}

Given the above arguments, we use the following two-step procedure to analyze the distribution: At the first stage, we compute the scale factor $a^2(g)$
in Eq.~(\ref{ascale}) for each of the $g$ values considered. Results for various system sizes up to $L=96$ are shown in Fig.~\ref{supfig1}. We can observe
that the curves for different system sizes cross each other around the transition point $g \approx 0.2175$. Based on closer inspection and extrapolations
of the crossing points, in the calculations of $I_q$ for for all $g$ and $L$ we fix $a(L)$ in Eq.~(\ref{mcondition}) at their values obtained at $g=0.21755$,
which is full consistent with the transition point $g_c=0.2175 \pm 0.0001$ extracted in the main paper.

\begin{figure}[t]
\includegraphics[width=84mm,clip]{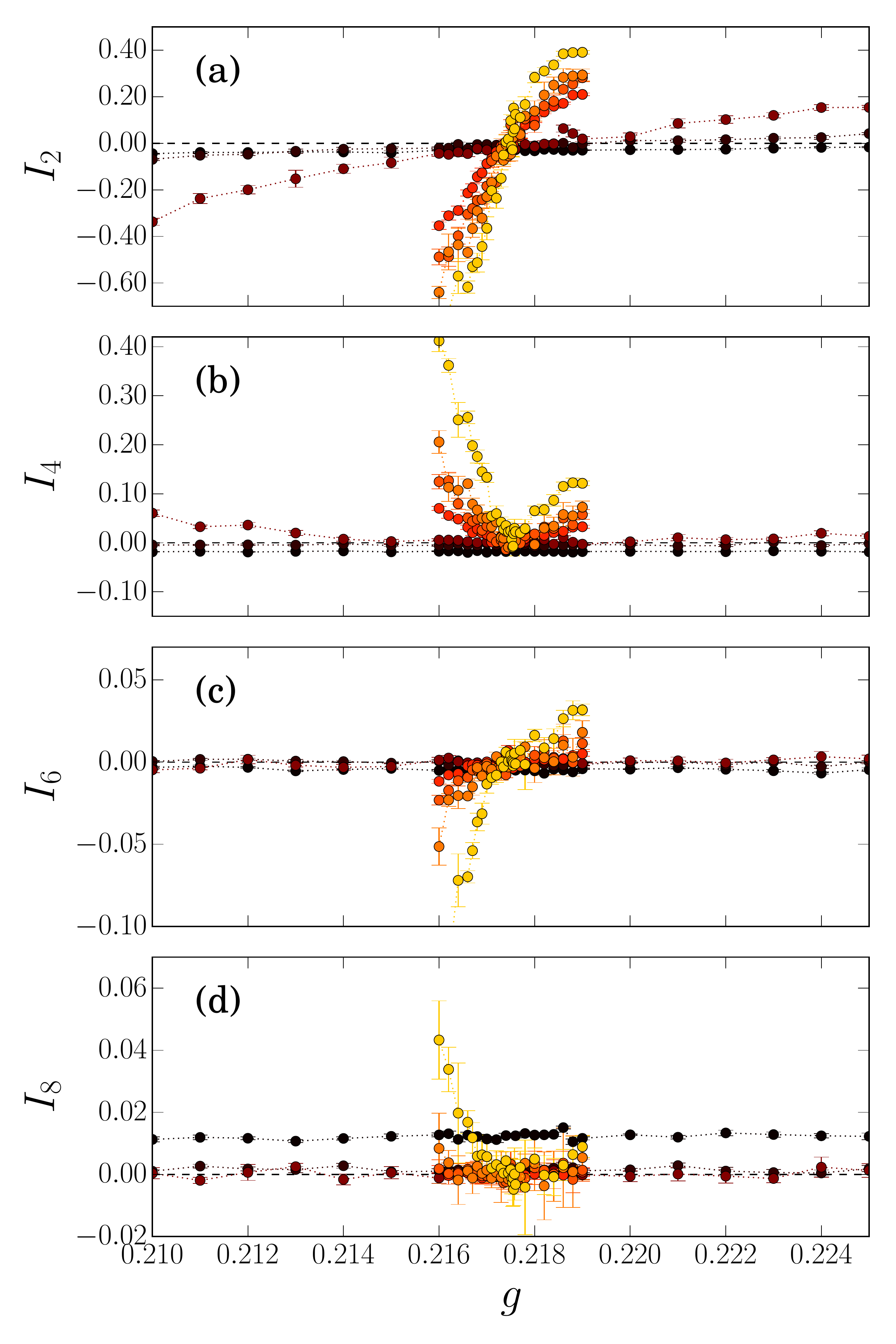}
\vskip-1mm
\caption{Tests of emergent O(4) symmetry, using the integrals $I_2$ (a), $I_4$ (b), $I_6$ (c), and $I_8$ (d). The value of the re-scaling parameter
$a(L)$ in Eq.~(\ref{ascale}) is held fixed at the value obtained in Fig.~\ref{supfig1}(c) at the point $g=0.21755$. The symbol colors correspond
to the system sizes as in Fig.~\ref{supfig1}.}
\label{supfig2}
\end{figure}

The results are shown in Fig.~\ref{supfig2}. Here we can see that the $I_q$ curves for odd $q$ go through zero at $g_c$, while for even $q$ they
exhibit minimums at $g_c$. For the larger system sizes, $I_4$ and $I_8$ have minimum values equal to $0$ to within statistical errors. Thus, we
conclude that there is an emergent O(4) symmetry at the transition point. Deviations from O(4) behvior are seen in $q_4$ and $I_8$ for the smallest
system sizes, which is expected if the symmetry is emergent upon increasing $L$. For the largest system sizes, up to $L=96$, no deviations from
O(4) symmetry can be detected at $g_c$ in any of the $I_q$ integrals.

\begin{figure}[t]
\includegraphics[width=84mm,clip]{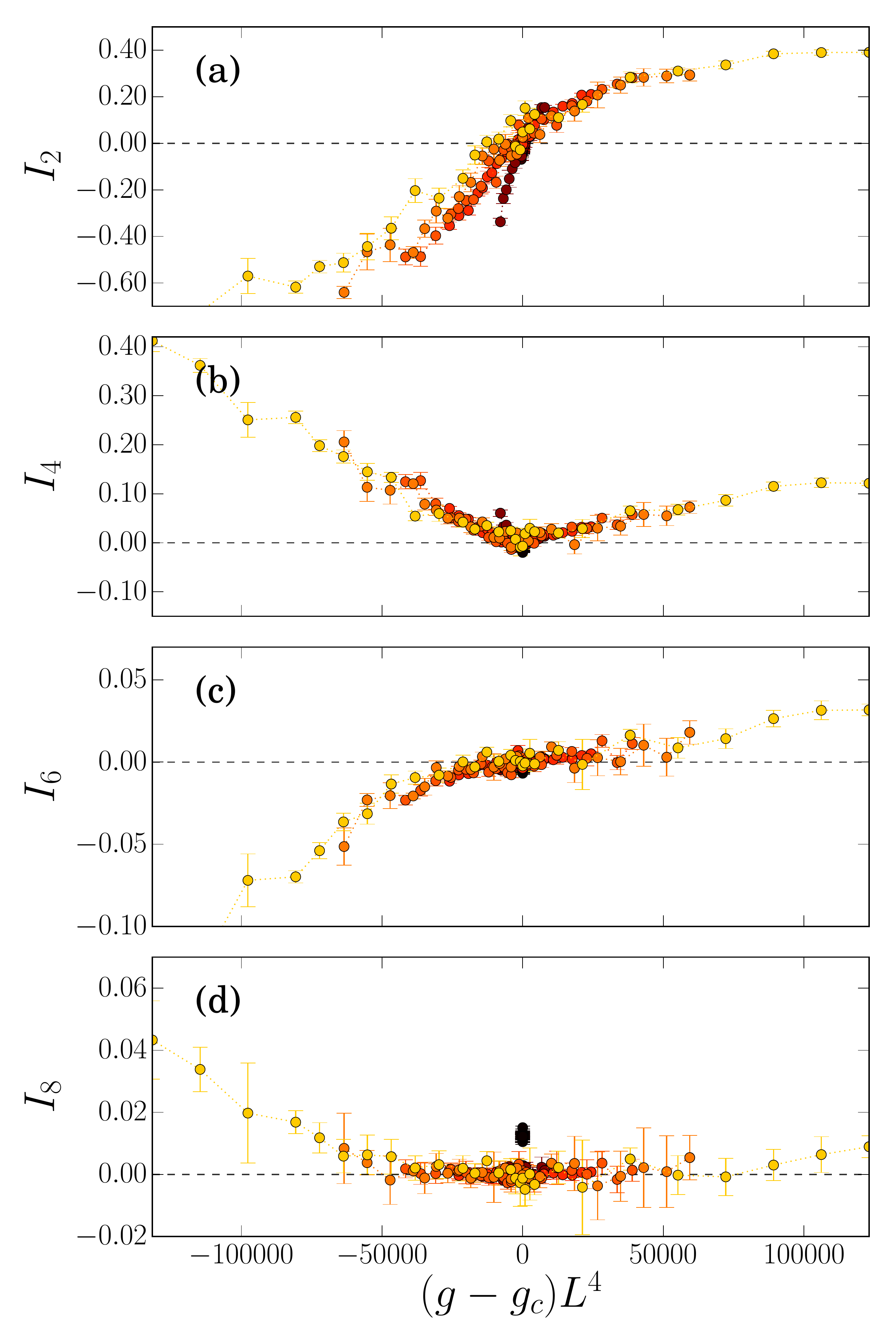}
\vskip-1mm
\caption{Tests of data collapse of $I_q(g,L)$ based on Eq.~(\ref{Eq:Collapse}). The data points are the same
as in Fig.~\ref{supfig2}, but the coupling has been rescaled by a factor $L^4$ relative to the critical point, for which we
here use $g_c=0.21755$, which is within the statistical errors of the value determined in the main paper.}
\label{SupFig:Collapse}
\end{figure}

We next carry out a data-collapse procedure with the data in Fig.~\ref{supfig2}. At a classical first-order phase transition, the exponent $\nu$ relevant for
finite-size scaling slightly away from the phase transition takes the trivial value $\nu=1/d$ \cite{Vollmayr_ZPB_1993,Iino_ArX_2017}. In analogy with the
case of quantum-critical points, one would expect that $d$ at a quantum first-order phase transition should be replaced by $d+z$, $z$ being the dynamic exponent
that relates time and length scales. In the case at hand, one would expect $z=2$, which corresponds to the scaling of the finite-size excitation gap,
$\epsilon  \propto L^{-2}$, in a 2D O(N) system. The time scale $\epsilon^{-1}$ corresponding to this energy is exactly that of the angular fluctuations
of the order parameter, as in Anderson's analogy between the finite-size excitations of an AFM and quantum rotor states (as studied explicitly
in QMC calculations in Ref.~\cite{Phil17}). Thus, we expect $1/\nu=z+2=4$ to be
the relevant scaling exponent describing the rotation of the O(4) vector as the transition point is traversed, and we attempt to describe the data by
the finite-size scaling form
\begin{equation}
I_q(g, L) = f_q [(g - g_c) L^4].
\label{Eq:Collapse}
\end{equation}
In Fig. \ref{SupFig:Collapse} we re-graph the data from Fig.~\ref{supfig2} versus $(g-g_c)L^4$. We indeed observe very good collapse of the data,
except for the smallest systems. These results lend additional support to an emergent O(4) symmetry.

It should be noted here that we did not see clearly that $1/\nu_{xy}$ and $1/\nu_{z}$ extrapolate to $4$ in Fig.~\ref{jqresults}(d). There could be
two reasons for this. First, as we mentiond also in Sec.~1 above, we have neglected corrections to the assumed leading linear scaling forms in
Fig.~\ref{jqresults}(d). Second, since the excitation gap at the transition should scale as $1/L^2$, there may be some effects of using $1/T \propto L$,
instead of $1/T \propto L^2$, in the SSE simulations. This would not affect any of the other extrapolated quantities in Fig.~\ref{jqresults}, as it is
still true that $T \to 0$ when $L\to \infty$ and the O(4) coexistence state should have an exponentially large correlation length $\xi(T)$ as $T\to 0$. i.e.,
$\xi(T) \gg L$ when $T \propto 1/L$. The projector QMC simulations used here to generate the order-parameter simulations are also fully consistent with
coexisting long-range orders and the value of the transition point extracted based on SSE simulations. As we have seen above, the quantitative scaling
analysis of the symmetry properties gives further support to the emergent O(4) symmetry in the complete absence of any finite-$T$ effects. 

Along with a finite order parameter at the transition, the O(4) symmetry projected down to two components also implies a flat radial distribution between $0$
and the radius $R$ of the sphere. As we pointed out in the main text and demonstrated in Fig.~\ref{pmdist}, the not completely flat behavior close to the rim
observed in the CBJQ histogram can be explained by finite-size fluctuations of the radius, which should vanish only in the limit $L \to \infty$.  Furthermore,
since the O(3) symmetry between the three components of the AFM order parameter is explicitly enforced by the Hamiltonian and is also not violated in any
way in the simulations, the demonstration of O(2) symmetry in the distribution $P(m_z,m_p)$ immediately also implies O(4) symmetry at the AFM-PSS
coexistence point. 

Thus, we have shown here that for the largest system size available, $L=96$, the CBJQ model has a point at which its combined AFM and PSS order parameters
exhibit O(4) symmetry to a high degree, with any potential violation too small to be detectable within the rather small error bars of our results. For the
smallest system sizes we do see some deviations from perfect symmetry, which is expected when the symmetry is not present in the hamiltonian but emerges
as the system size increases.
\vskip2mm

\begin{figure}[t]
\includegraphics[width=84mm,clip]{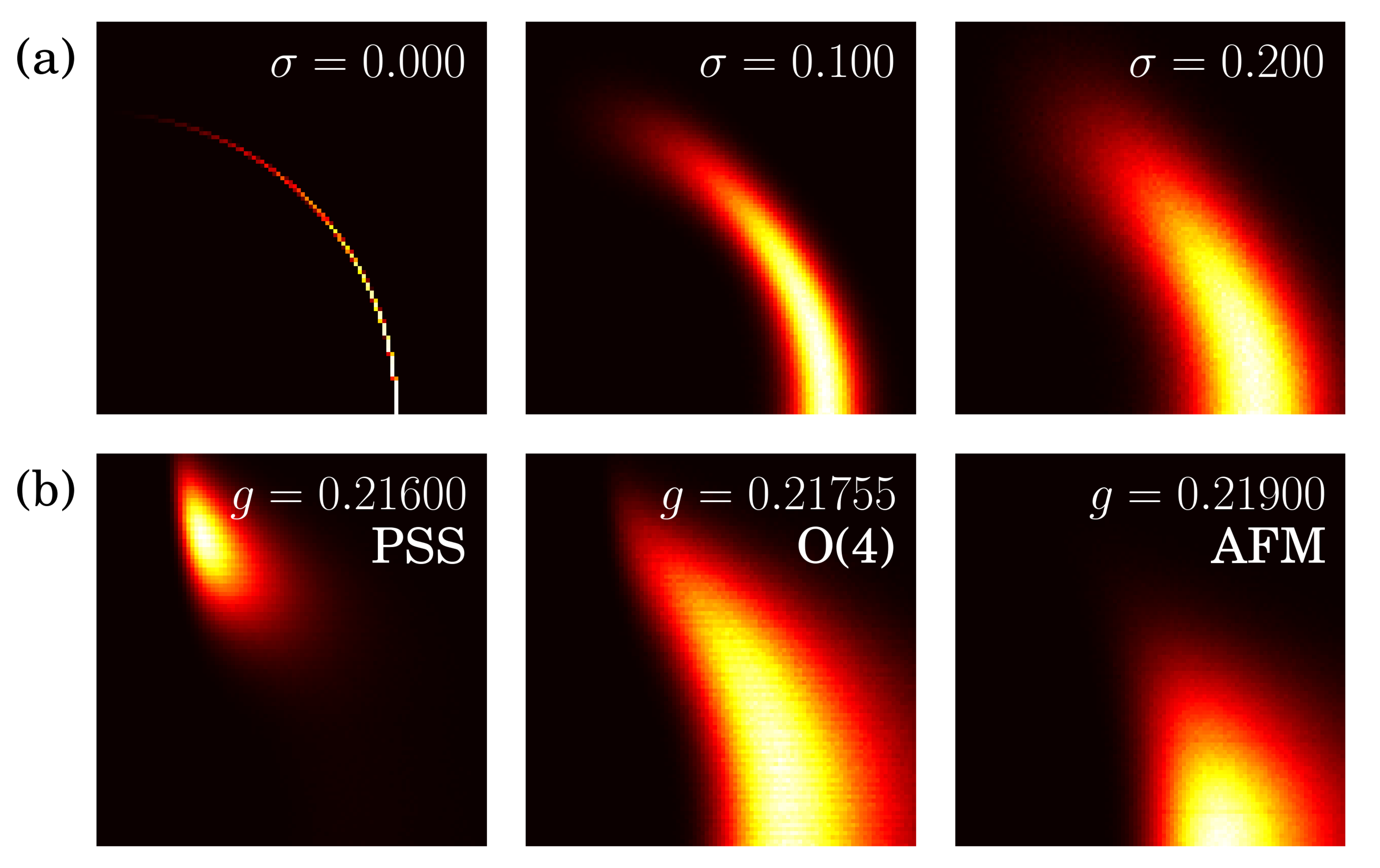}
\vskip-1mm
\caption{Test of emergent O(4) symmetry of the CBJQ involving all four order parameter components. 
The $x$ axis represents the magnitude of the total AFM order
parameter $m_s$, defined in Eq.~(\ref{msdef}), and the vertical axis the PSS order parameter $m_p$. Only the quadrant with all positive values is
shown. Panels (a) are for the case of a perfect O(4) sphere with radius $R=1$ and variance $\sigma^2$, sampled using the algorithm in
Ref.~\cite{Muller_ACM_1959}. Panels (b) show valence-bond projector QMC results for the CBJQ model at three values of $g$; inside the PSS phase,
close to the transition point with emergent O(4) symmetry, and inside the AFM phase.}
\label{supfig3}
\end{figure}

{\it Tests with four components.}---We complement the above analysis of two out of the four components of the putative O(4) vector with a test where all four
components are used, projected down to two dimensions by using the magnitude of the full O(3) AFM order parameter in Eq.~(\ref{msdef}) and the PSS order
parameter, i.e., the distribution $P(m_s,m_p)$. We carry out a process similar to the one discussed above to put the overall lengths of the AFM and PSS
components on equal footing.

For an ideal O(4) sphere with fixed $R$ projected down to two dimensions in this manner, the distribution $P(m_s,m_p)$ has the shape of arc of infinitesimal
thickness and radius $R$, with the density varying proportionally to $m_s^2$ along the arc, due to the different contents (number of components of the
4-dimensional vector) of the two dimensions. Fig.~\ref{supfig3}(a) shows the distribution for three different values of the standard deviation $\sigma$
of the fluctuating radius about the mean value $R=1$. In the case of the CBJQ model, as shown in Fig.~\ref{supfig3}(b), there is indeed very little weight
close to the y-axis as expected. As we go from the PSS state to the AFM state the weight shifts clockwise from large $y$ ($m_p$) values down toward the
$x$-axis (large $m_s$). At the transition point we see a distribution very similar to the O(4) sphere with $\sigma \approx 0.20$ 

It should be noted that $m_s^2$ in the valence-bond basis is obtained from the transition graph as a sum of squared loop lengths, and this corresponds to a 
sum over $2^{n_l}$ spin configurations in the basis of $S^z$ spins, $n_l$ being the number of loops (each loop having two compatible staggered spin 
configurations). This implicit averaging over points on the putative O(4) sphere may cause some additional smearing in $P(m_s,m_p)$, beyond just the 
projection down to two dimensions and the fluctuations of the radius associated with finite system size. The somewhat larger $\sigma$ required to match the O(4) 
sphere in Fig.~\ref{supfig3} than what was needed in the case of $P(m_z,m_p)$ in Fig.~\ref{pmdist} likely reflects this effect. In addition, for finite 
system size, the loop estimator for $m_s$ has a strict lower bound $\propto L^{-2}$, with a de facto large prefactor, and this also seems to cause some 
visible deviations from the O(4) sphere results at the left tip of the distribution. For these reasons, we believe that the $P(m_z,m_p)$ distribution 
is better for quantitatively characterizing the degree of symmetry.
\vskip3mm

\subsection{3. The CBJQ model with $Q_A \not= Q_B$}

\begin{figure}[t]
\includegraphics[width=84mm,clip]{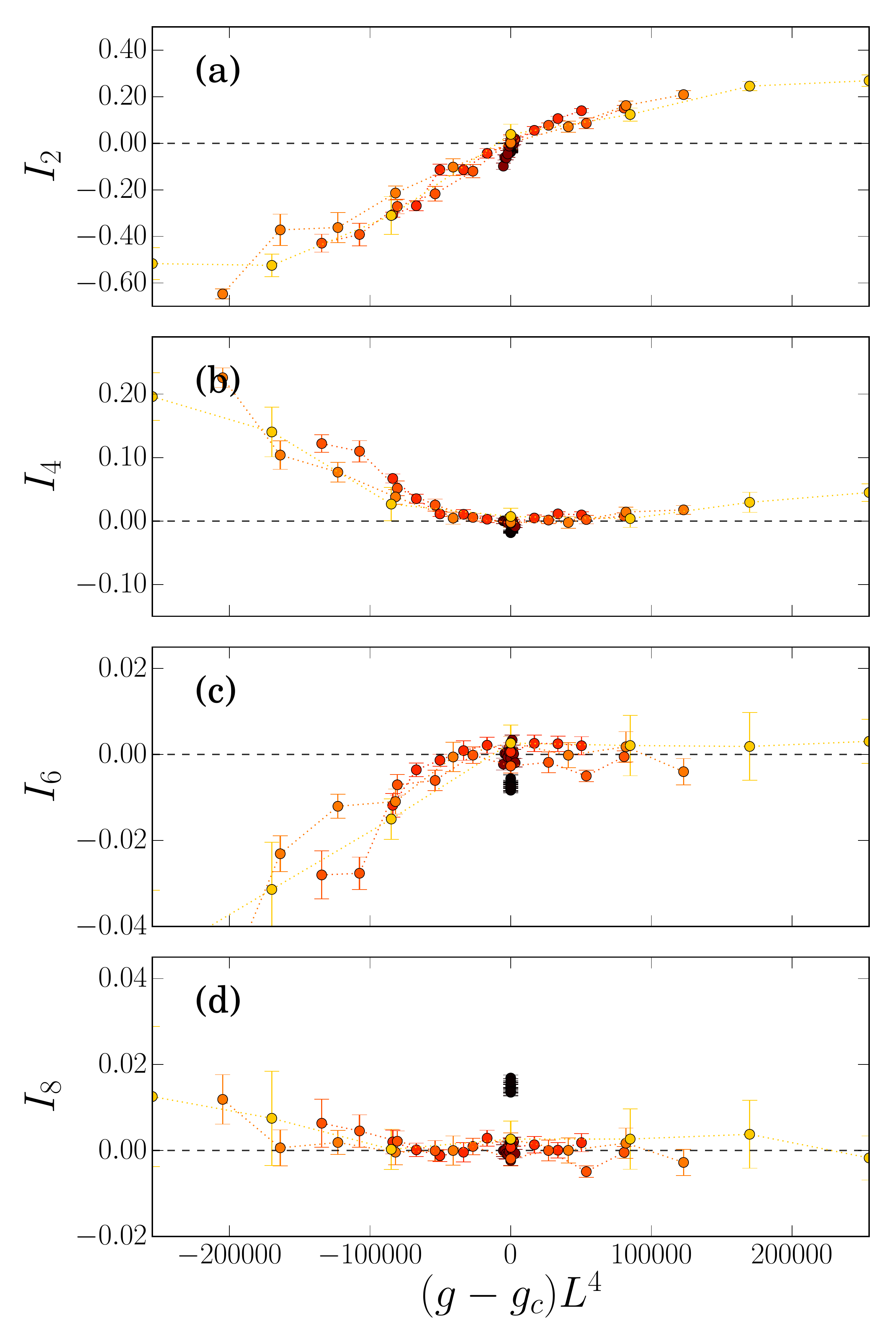}
\vskip-1mm
\caption{Data collapse of the integrals $I_q(g,L)$ for the extended CBJQ model with $Q_B=Q_A/2$, based on Eq.~(\ref{Eq:Collapse}).
The system sizes are up to $L=96$, with the symbols color-coded as in Fig.~\ref{supfig1}.}
\label{SupFig:Collapse2}
\end{figure}

One might wonder whether the CBJQ model could have an accidental emergent symmetry, i.e., that a generically present symmetry-breaking perturbation at the
transition point could be absent, or extremely small, in this particular model. In order to exclude such a fine-tuning scenario, we here deform the CBJQ model
in a significant way by introducing four-spin coupling of strenth $Q_B$ on the plaquettes without $Q$-couplings in Eq.~(\ref{jqham}) and Fig.~\ref{figmodels}.
We refer to the previously present plaquette couplings as $Q_A$. Calling the two sets of plaquettes $\Box_A$ and $\Box_B$, the Hamiltonian of
this extended CBJQ model is
\begin{eqnarray}
H = -J \sum_{\langle ij \rangle} && P_{ij} -Q_A \hskip-2mm \sum_{ijkl \in \Box_A} \hskip-1mm (P_{ij} P_{kl} + P_{ik} P_{jl}) \nonumber \\
&& -Q_B \hskip-2mm \sum_{ijkl \in \Box_B} \hskip-1mm (P_{ij} P_{kl} + P_{ik} P_{jl}).
\label{jqhamab}
\end{eqnarray}
When $Q_A=Q_B$, this model becomes the ordinary $J$-$Q$ model on the square lattice, which hosts a four-fold degenerate columnar VBS and where no
convincing signs of a first-order transition between it and the AFM has been detected. The non-magnetic state in the extended CBJQ model is likely
a two-fold degenerate PSS for any $Q_A \not= Q_B$, though this is not completely clear because the checkerboard deformation of the square-lattice
system is still also compatible with a four-fold degenerate columnar VBS. If indeed we have the PSS for all $Q_A\not=Q_B$, then for $Q_A$ very close
to $Q_B$ there will be interesting cross-over behaviors as the system size is increased, from an almost $Z_4$ symmetric columnar VBS to a $Z_2$ symmetric
PSS. However, the first-order behavior we have found here for $Q_B=0$ may in principle end at a tricritical point at some $0 < (Q_B/Q_A)_t < 1$, with
a continuous AFM--PSS transition obtaining for $(Q_B/Q_A)_t < Q_B/Q_A < 1$. In a future study, we plan to investigate these and other issues for the extended
CBJQ model in its full range of the ratio $Q_B/Q_A$. 

Here we only consider $Q_B/Q_A=1/2$, for which we find that the transition is still first-order and the nonmagnetic state is indeed the PSS. We
focus on the histogram approach for detecting emergent O(4) symmetry. The added interaction is sufficiently strong so that an accidental emergent
symmetry cannot be the explanation for our finding for $Q_B/Q_A=0$ if we also find O(4) symmetry when $Q_B/Q_A=1/2$---unless this whole class of
extended CMJQ models is automatically fine-tuned, which by itself would be remarkable and an indication that symmetry perturbations can be
broadly avoided (likely beyond the CBJQ models) under previously overlooked circumstances.

We proceed using the same methods as previously, here up to system size $L=96$. We find $g_c = 0.139 \pm 0.001$ and the length of the O(4) vector with the
same normalization as in the previous section is about 10\% of the maximum value,  or $40\%$ of the value found when $Q_B=0$. Scaled results for the
integrals $I_q$ are shown in Fig.~\ref{SupFig:Collapse2}. Here we observe the same kind of behavior as in Fig.~\ref{SupFig:Collapse}, with good data
collapse and no discernible deviations from $0$ of the $I_q$ values at $g_c$. Thus, we conclude that the emergent O(4) symmetry is not due to some
accidental fine-tuning but an intrinsic feature of the CBJQ class of models.

Comparing Figs.~\ref{SupFig:Collapse} and \ref{SupFig:Collapse2}, we see that the scaled integrals are close to zero over a wider range of $(g-g_c)L^4$ when
the first-order transition is weker, i.e., for $Q_B/Q_A=1/2$. This is natural in light of the larger fluctuations as the discontinuities of the order
parameters weaken. The behavior also points to an important role of a quantum-critical point that may be reachable as $Q_B/Q_A$ is further increased
(but not up to $Q_B/Q_A=1$, where the non-magnetic state is different, a columnar VBS breaking $Z_4$ symmetry). The emergent symmetry may ultimately
be connected to some extension of the DQCP scenario, as also mentioned in the main paper.

These results of course do not prove that the emergent symmetry exists up to infinite length scale. In the scenario of an approximate symmetry at
the first-order transition being due to a nearby critical point with exact emergent symmetry, one would expect the length scale up to which the
apprpximate symmetry applies to be larger when the order parameter is smaller. Therefore, it would clearly be useful to also study a model where the
coexisting order parameters are larger than for the case $Q_B=0$. However, so far we have not found any way to make the transition more strongly first-order
within the CBJQ  models.


\begin{thebibliography}{99}
\bibitem{so4note}
Our calculations cannot distinguish between O(4) and SO(4) symmetry, since we only test for the rotational symmetry
common to both.
\end{thebibliography}
\end{document}